\documentclass{article}

\input{tcilatex}

\begin{document}

\subsection{\ \ \ \ \ \ \ \ \ \ \ \ \ \ \ \ \ \ \ \ \ \ \ \ \ \ \ \ \ \ The
Enigma of the Mass.}

\ \ \ \ \ \ \ \ \ \ \ \ \ \ \ \ \ \ \ \ \ \ \ \ \ \ \ \ \ \ \ \ \ \ \ \ \ \
\ \ \ \ \ \ \ \ \ \ \ \ \textbf{V.G. Plekhanov}

\textbf{\ \ \ \ \ \ \ \ \ \ \ \ \ \ \ \ \ Computer Science College, Erika
Street 7a, Tallinn, 10416, Estonia}

\bigskip

\textbf{\bigskip Abstract.}

The different manifestations of the mass effects in the microphysics
(isotope effect) are presented for the first time. The bright effects
observe in \ all branches of physics: nuclear, atomic, and molecular as well
as solid state physics. Charge symmetry breaking in the strong interaction
occurs because of the difference between the \textbf{masses }of the up and
down quarks. At present the Standard Model can't explain the observed mass
pattern (M$_{n}$, M$_{p}$, m$_{u}$, m$_{d}$ etc.) \ and their hierarchy. The
last one doesn't permit us to find the origin of the isotope effect. The
origin of the mass of the matter will be clarified when the mechanism of
chiral symmetry breaking in QCD is established.

\bigskip

Mass is a one of the fundamental properties of matter. It relates to
classical as well as modern physics (quantum mechanics or modern theory of
gravitation (see, e.g. [1]). Although the physical meaning of \ mass was
discovered by Einstein more than a century ago, when he introduced in
physics the concept of rest energy (E$_{0}$) [2], the concept of mass still
doesn't have strict mathematical determination. Indeed, according to the
notion of the relativistical physics (see, e.g. [3]) mass is determined by
the next expression

m$^{2}$ = $\frac{\text{E}^{2}}{\text{c}^{4}}$ - $\frac{\overrightarrow{\text{%
p}^{2}}}{\text{c}^{2}}$ \ \ \ \ \ \ \ \ \ \ \ \ \ \ \ \ \ \ \ \ \ \ (1).

And in the case of resting body ($\overrightarrow{\text{p}}$ = 0) we have

m = $\frac{\text{E}_{0}}{\text{c}^{2}}$ \ \ \ \ \ \ \ \ \ \ \ \ \ \ \ \ \ \
\ \ \ \ \ \ \ \ \ \ \ \ \ (2).

From equation (2) it can be seen that the mass is proportioned to the rest
energy. If we put c = 1, in that case we see that the mass of body equals
its rest energy. The mass of a body is not a constant, it varies with
changes in its energy. Namely, rest energy "slumbering" in massive bodies
partly is released in chemical and especially nuclear reactions. In spite \
of equivalence of the mass of the body and rest energy, especially nuclear
physics and physics of elementary particles, the task of mass has not been
solved. Until present time the spectrum of the discrete hierarchy of
elementary particles mass \ hasn't had a successful theoretical explantion
[4,6]. \ As is well-known \ on the boundary of the 19 and 20 centuries there
was an opinion that the mass of the electron has the electromagnetic origin
[1,4]. However, later investigations showed that the electromagnetic part of
the mass of the electron has a small contribution to its full mass [3].
Nevertheless, the modern view connects the origin of the mass with nonlocal
gravitational fields, which nature is due to electromagnetic interaction [8
-11].This conclusion \ reflects those fact, that the space between separated
particles in essence isn't empty, it is filled with the material medium -
the physical fields. The space inside the atom is filled with
electromagnetic field, and inside nucleus - more densier and stronger field
which is called sometimes meson one.

The present letter is devoted to the elucidation of the origin of mass, so
far as only its nature closely connected with the origin of the isotope
effect, the experimental manifestation of which more persuasively testified
in the last fifty years\ in all branches of physics (nuclear, atomic,
molecular as well as solid state (see, e.g. reviews [12-14])). On the other
hand it is necessary to underline that only isotope effect is a direct
manifestation of the mass effect in microphysics. It should be added that
the direct measurments of the energy of zero-point vibrations owing to
isotope effect in solids shows the good agreement of the experimental values
with the results of the calculation of quantum electrodynamics in solids
[13, 14].

Below we describe shortly the manifestations of the isotope effect in
molecular as well as solid state physics (more details see [14]). The
discovery [15] of the new fullerene allotropes of carbon, exemplified by C$%
_{60}$ and soon followed by an efficient method for their synthesis [14],
led to a burst of theoretical and experimental activity on their physical
properties. Much of this activity concentrated on the vibrational properties
of C$_{60}$ and their elucidation by Raman scattering [15]. Comparison \
between theory and experiment was greatly simplified by the high symmetry (I$%
_{h}$), resulting in only ten Raman active modes for the isolated molecule
and the relative weakness of solid state effect [15], causing the
crystalline C$_{60}$ (c - C$_{60}$) Raman spectrum at low resolution to
deviate only slightly from that expected for the isolated molecule [15].
Since the natural abundance of $^{13}$C is 1.11\% (see, e.g. [12]), almost
half of all \ C$_{60}$ molecules made from natural graphite contain one or
more $^{13}$C isotopes. If the squared frequency of a vibrational mode in a C%
$_{60}$ molecule with n$^{13}$C isotopes is written as a series $\ \omega
^{2}$ = $\omega _{(0)}^{2}$ + $\omega _{(1)}^{2}$ + $\omega _{(2)}^{2}$ + $%
\omega _{(3)}^{2}$ + ...... in the mass perturbation (where $\omega _{\left(
0\right) }$ is an eigenmode frequency in a C$_{60}$ molecule with 60 $^{12}$%
C atoms), nondegenerate perturbation theory predicts for the two totally
symmetric A$_{g}$ modes a first - order correction given

$\frac{\omega _{(1)}^{2}}{\omega _{(0)}^{2}}$ = - $\frac{\text{n}}{\text{720}%
}$. \ \ \ \ \ \ \ \ \ \ \ \ \ \ \ \ \ \ \ \ \ \ \ \ \ \ \ \ \ (3)

This remarkable result, independent of the relative position of the isotopes
within the molecule and equally independent of the unperturbed eigenvector,
is a direct consequence of the equivalence of all carbon atoms in
icosahedral C$_{60}$. To the same order of accuracy within nondegenerate
perturbation theory, the Raman polarizability derivatives corresponding to
the perturbed modes are equal to their unperturbed counterparts, since the
mode eigenvectors remain unchanged. These results lead to the following
conclusion [15]: The A$_{g}$ Raman spectrum from a set of noninteracting C$%
_{60}$ molecules will mimic their mass spectrum if the isotope effect on
these vibrations can be described in terms of first - order nondegenerate
perturbation theory. It is no means obvious that C$_{60}$ will meet the
requirements for the validity of this simple theorem. A nondegenerate
perturbation expansion is only valid if the A$_{g}$mode is sufficiently
isolated in frequency from its neighboring modes. Such isolation is not, of
course, required by symmetry. Even if a perturbation expansion \ converges,
there is no a priori reason \ why second - and higher - order correction to
Eq. (3) should be negligible. As was shown in cited paper the experimental
Raman spectrum (see below) of C$_{60}$ does agree with the prediction of Eq.
(3). Moreover, as was shown in quoted paper, experiments with isotopically
enriched \ samples display the striking correlation between mass and Raman
spectra predicted by the above simple theorem. Fig. 1 shows a high -
resolution Raman spectrum at 30 K in an energy range close to the high -
energy pentagonal - pinch A$_{g}$(2) vibration according to [15]. Three
peaks are resolved, with integrated intensity of 1.00; 0.95; and 0.35
relative to the strongest peak. The insert of this figure shows the
evolution of this spectrum as the sample is heated. The peaks cannot be
resolved beyond the melting temperature of CS$_{2}$ at 150 K. The
theoretical fit yields a separation of 0.98 $\pm $ 0.01 cm$^{-1}$ between
two main peaks and 1.02 $\pm $ 0.02 cm$^{-1}$ between the second and third
peaks. The fit also yields full widths at half maximum (FDWHM) of 0.64; 0.70
and 0.90 cm$^{-1}$, respectively. The mass spectrum of this solution shows
three strong peaks (Fig. 1$^{b}$) corresponding to mass numbers 720; 721 and
722, with intensities of 1.00; 0.67 and 0.22 respectively as predicted from
the known isotopic abundance of $^{13}$C. The authors [15] assign the
highest - energy peak at 1471 cm$^{-1}$ to the A$_{g}$(2) mode of
isotopically pure C$_{60}$ (60 $^{12}$C atoms). The second peak at 1470 cm$%
^{-1}$ is assigned to C$_{60}$ molecules with one $^{13}$C isotope, and the
third peak at 1469 cm$^{-1}$ to C$_{60}$ molecules with \ two $^{13}$C
isotopes. The separation \ between the peaks is in excellent agreement with
the prediction from Eq. (3), which gives 1.02 cm$^{-1}$. In addition, the
width of the Raman peak at 1469 cm$^{-1}$, assigned to a C$_{60}$ molecule
with two $^{13}$C atoms, is only 30 \% larger than the width of the other
peaks. This is consistent with the prediction of Eq. (3) too, that the
frequency of the mode will be independent of the relative position of the $%
^{13}$C isotopes within the molecule. The relative intensity between two
isotope and one isotope Raman lines agrees well with the mass spectrum
ratios. Concluding this part we stress that the Raman spectra of C$_{60}$
molecules show remarkable correlation with their mass spectra. Thus the
study of isotope - related shift offers a sensitive means to probe the
vibrational dynamics of C$_{60}$.

Next examples of the dependence of the exciton spectra in solds on the
isotope effect demonstrate below. Isotopic substitution only affects the
wavefunction of phonons; therefore, the energy values of electron levels in
the Schr\"{o}dinger equation ought to have remained the same. This, however,
is not so, since isotopic substitution modifies not only the phonon
spectrum, but also the constant of electron-phonon interaction (see [12]).
It is for this reason that the energy values of purely electron transition
in molecules of hydride and deuteride are found to be different. This effect
is even more prominent when we are dealing with a solid [16].
Intercomparison of absorption spectra for thin films of LiH and LiD at room
temperature revealed that the longwave maximum (as we know now, the exciton
peak ) moves 64.5 meV towards the shorter wavelengths when H is replaced
with D [17].

The mirror reflection spectra of mixed and pure LiD crystals cleaved in
liquid helium are presented in Fig. 2. For comparison, on the same diagram
we have also plotted the reflection spectrum of LiH crystals with clean
surface. All spectra have been measured with the same apparatus under the
same conditions. As the deuterium concentration increases, the long-wave
maximum broadens and shifts towards the shorter wavelengths. As can clearly
be seen in Fig. 2, all spectra exhibit a similar long-wave structure. This
circumstance allows us to attribute this structure to the excitation of the
ground (Is) and the first excited (2s) exciton states. The energy values of
exciton maxima for pure and mixed crystals at 2 K are presented in Table 22
of ref. [12]. The binding energies of excitons E$_{\text{b}}$, calculated by
the hydrogen-like formula, and the energies of interband transitions E$_{%
\text{g}}$ are also given in Table 22.

Going back to Fig. 2, it is hard to miss the growth of $\Delta _{\text{12}}$%
, which in the hydrogen-like model causes an increase of the exciton Rydberg
with the replacement of isotopes . When hydrogen is completely replaced with
deuterium, the exciton Rydberg (in the Wannier-Mott model) increases by 20\%
from 40 to 50 meV, whereas E$_{\text{g}}$ exhibits a 2\% increase, and at 2 $%
\div $ 4.2 K is $\Delta $E$_{\text{g}}$ = 103 meV. This quantity depends on
the temperature, and at room temperature is 73 meV, which agrees well enough
with $\Delta $E$_{\text{g}}$ = 64.5 meV as found in the paper of Kapustinsky
et al. [17]. The single-mode nature of exciton reflection spectra of mixed
crystals LiH$_{\text{x}}$D$_{\text{1-x}}$ agrees qualitatively with the
results obtained with the virtual crystal model (see e.g. Elliott et al.
[18]; Onodera and Toyozawa [19]), being at the same time its extreme
realization, since the difference between ionization potentials ($\Delta
\zeta $) for this compound is zero. According to the virtual crystal model, $%
\Delta \zeta $ = 0 implies that $\Delta $E$_{\text{g}}$ = 0, which is in
contradiction with the experimental results for LiH$_{\text{x}}$D$_{\text{1}%
} $-$_{\text{x}}$ crystals. The change in E$_{\text{g}}$ caused by isotopic
substitution has been observed for many broad-gap and narrow-gap
semiconductor compounds (see also [12]).

All of these results are documented in Table 22 of Ref.[12], where the
variation of E$_{\text{g}}$, E$_{\text{b}}$, are shown at the isotope
effect. We should highlighted here that the most prominent isotope effect is
observed in LiH crystals, where the dependence of E$_{\text{b}}$ = f (C$_{%
\text{H}}$) is also observed and investigated. To end this section, let us
note that E$_{\text{g}}$ decreases by 97 cm$^{\text{-1}}$ when $^{\text{7}}$%
Li is replaced with $^{\text{6}}$Li.

Detailed investigations of the exciton reflectance spectrum in CdS crystals
were done by Zhang et al. [20]. Zhang et al. studied only the effects of Cd
substitutions, and were able to explain the observed shifts in the band gap
energies, together with the overall temperature dependence of the band gap
energies in terms of a two-oscillator model provided that they interpreted
the energy shifts of the bound excitons and n = 1 polaritons as a function
of average S mass reported earlier by Kreingol'd et al. [21] as shifts in
the band gap energies. However, Kreingol'd et al. [21] had interpreted these
shifts as resulting from isotopic shifts of the free exciton binding
energies , and not the band gap energies, based on their observation of
different energy shifts of features which they identified as the n = 2 free
exciton states (for details see [21]). The observations and interpretations,
according Meyer at al. [22], presented by Kreingol'd et al. [21] are
difficult to understand, since on the one hand a significant band gap shift
as a function of the S mass is expected , whereas it is difficult to
understand the origin of the relatively huge change in the free exciton
binding energies which they claimed. Very recently Meyer et al. [22]
reexamine the optical spectra of CdS as function of average S mass, using
samples grown with natural Cd and either natural S ($\sim $ 95\% $^{32}$S),
or highly enriched (99\% $^{34}$S). These author observed shifts of the
bound excitons and the n = 1 free exciton edges consistent with those
reported by Kreingol'd et al. [21], but, contrary to their results, Meyer et
al. observed essentially identical shifts of the free exciton excited
states, as seen in both reflection and luminescence spectroscopy. The
reflectivity and photoluminescence spectra in polarized light ($%
\overrightarrow{E}$ $\bot $ $\overrightarrow{C}$) over the A and B exciton
energy regions for the two samples depicted on the Fig. 3. For the $%
\overrightarrow{E}$ $\bot $ $\overrightarrow{C}$ polarization used in Fig. 3
both A and B excitons have allowed transitions, and therefore reflectivity
signatures. Fig. 3 reveals both reflectivity signatures of the n = 2 and 3
states of the A exciton as well that of the n = 2 state of the B exciton.

In Table 18 of Ref. [14] the results of Meyer et al. summarized the energy
differences $\Delta $E = E (Cd$^{34}$S) - E (Cd$^{nat}$S), of a large number
of bound exciton and free exciton transitions, measured using
photoluminescence, absorption, and reflectivity spectroscopy, in CdS made
from natural S (Cd$^{nat}$S, 95\% $^{32}$S) and from highly isotopically
enriched $^{34}$S (Cd$^{34}$S, 99\% $^{34}$S). As we can see from Fig. 3,
all of the observed shifts are consistent with a single value, 10.8$\pm $0.2
cm$^{-1}$. Several of the donor bound exciton photoluminescence transitions,
which in paper [22] can be measured with high accuracy, reveal shifts which
differ from each other by more than the relevant uncertainties, although all
agree with the 10.8$\pm $0.2 cm$^{-1}$ average shift. These small
differences in the shift energies for donor bound exciton transitions may
reflect a small isotopic dependence of the donor binding energy in CdS (see,
also [12]). This value of 10.8$\pm $0.2 cm$^{-1}$ shift agrees well with the
value of 11.8 cm$^{-1}$ reported early by Kreingol'd et al. [21] for the B$%
_{n=1}$ transition, particularly when one takes into account \ the fact that
enriched $^{32}$S was used in that earlier study, whereas Meyer et al. have
used natural S in place of an isotopically enriched Cd$^{32}$S (for details
see [22]).

Authors [21] conclude that all of the observed shifts \ arise predominantly
from an isotopic dependence of the band gap energies, and that the
contribution from any isotopic dependence of the free exciton binding
energies is much smaller. On the basis of the observed temperature
dependencies of the excitonic transitions energies, together with a simple
two-oscillator model, Zhang et al. [20] earlier calculated such a
difference, predicting a shift with the S isotopic mass of 950 $\mu $eV/amu
for the A exciton and 724 $\mu $eV/amu for the B exciton. Reflectivity and
photoluminescence study of $^{nat}$Cd$^{32}$S and $^{nat}$Cd$^{34}$S
performed by Kreingol'd et al. [21] shows that for anion isotope
substitution the ground state (n = 1) energies of both A and B excitons have
a positive energy shifts with rate of $\partial $E/$\partial $M$_{S}$ = 740 $%
\mu $eV/amu. Results of Meyer et al. [22] are consistent with a shift of $%
\sim $710 $\mu $eV/amu for both A and B excitons. Finally, it is interesting
to note that the shift of the exciton energies with Cd mass is 56 $\mu $%
eV/amu [20], an order of magnitude less than found for the S mass (more
details see [12, 13]).

The brought examples clearly indicate mass dependence of the electron and
phonon states (see more details [14]) but on the other side it simply
underlines the primary importance in microphysics the difference of mass
between neutron (M$_{n}$) and proton (M$_{p}$). Really small difference in
their masses M$_{n}$ - M$_{p}$ = 1.2333317 MeV leads to the bright effects
in microphysics. According to the last data [9], the experimental
neutron-proton mass difference of M$_{n}$ - M$_{p}$ = 1.2333317 MeV is
received as estimated electromagnetic contribution M$_{n}$ - M$_{p}\mid ^{%
\text{em}}=$ -0.76 $\pm $ 0.30 MeV, and the remaining mass difference is
determined to a strong isospin breaking contribution of M$_{n}$ - M$_{p}\mid
^{\text{d-u}}$= 2.05 $\pm $ 0.30 MeV. In other words the last contribution
is a result of difference in mass of d - and u - quarks (see, also [10, 11]).

As we all know, the observed world - stars, planets, galaxy as well as
surrounding objects consist from the nuclei, neutrons, protons and
electrons. The mass of \ electrons has a small contribution to the total
mass ( less than 0.1\% (see, e.g. [1]). Therefore, that we knew that the
origin of the mass of the observed worlds needs to be elucidated the origin
of nuclear mass. As we know the nucleon consists from u - and d - quarks.
But the mass of u - and d - quarks is so small, that is their sum is a small
part of the nucleon mass (1 - 2 \% [6]). In modern physics of elementary
particles it is considered that the mass of nucleon arises from the
spontaneous breaking of a chiral symmetry in quantum chromodynamics (QCD)
[23] and may be expressed over vacuum condensate (see [5] and references
therein).This model has an approximate formula which expresses the mass of
nucleon over quarks condensate [5]

m = [-2(2$\pi $)$^{2}\langle $0$\mid \overline{\text{q}}$q$\mid $0$\rangle $]%
$^{1/3}$ \ \ \ \ \ \ \ \ \ \ \ \ \ \ \ \ \ \ \ \ \ \ \ \ (4),

where m is nucleon mass, $\langle $0$\mid \overline{\text{q}}$q$\mid $0$%
\rangle $ is quarks condensate, q is the field of u - \ or d - quarks. The
chiral symmetry in QCD tresult in the expression for the quarks condensate
(so called Gell - Mann - Oakes - Renner formula [24])

\ $\langle $0$\mid \overline{\text{q}}$q$\mid $0$\rangle $ = -$\frac{\text{1}%
}{\text{2}}\frac{\text{m}_{\pi }\text{ f}_{\pi }}{\text{m}_{u}\text{ + m}_{%
\text{d}}}$ \ \ \ \ \ \ \ \ \ \ \ \ \ \ \ \ \ \ \ \ \ \ \ \ \ \ \ \ \ \ (5).

Here m$_{\pi }$ and f$_{\pi }$ are the mass and decay constant of $\pi $ -
meson. The defined value of quarks condensate on the ground of $\tau $ -
decay [5,6] equals

$\langle $0$\mid \overline{\text{q}}$q$\mid $0$\rangle $ = - (254 MeV)$^{3}$ 
$\pm $ 10 \%. \ \ \ \ \ \ \ \ \ (6)

Put the last value into the expression (4) it gives the nuclon's mass m =
1.08 GeV, when the experimental value of nucleon's mass equals m = 0.94 MeV.
From comparison of these values \ we see that the difference between
experimental value of m and theoretical estimation is 0.15 GeV, that
surpasses the experimental value of the difference M$_{n}$ - M$_{p}$ =
1.2333317 MeV much order. The last one means that in such model (as well as
in the model of constituent quarks) we have neither the mass difference of
the nucleons nor its number in nuclei and, consequently, isotope effect. \
But the experimental manifestations of the isotope effect was demonstrated
above \ in the different branches of microphysics. Considering the quarks
structure of nucleon (the wavefunction of the neutron is udd, and for proton
one is uud) that is the quark strucure indicates the different construction
of the neutron and proton, but this model doesn't quantative describe the
mass of nucleons.

Thus, the origin of the isotope effect is closely connected with the
different origin of u - and d - quarks and with solution the spectrum and
hierarhy of the elementary particles mass and more common with the solution
of the nature of mass (see, also [25]).

\bigskip

\textbf{Acknowledgements.} I deeply thank to Prof. B.L. Ioffe for the
enlighting discussion on the origin of mass, and Prof. P. Sneider for
improving my English.

\bigskip

\textbf{Figure captions.}

Fig. 1. a - unpolarized Raman spectrum in the frequency region of the
pentagonal - pinch mode, for a frozen sample of nonisotopically enriched C$%
_{60}$ in CS$_{2}$ at 30 K. The points are the experimental data, and the
solid curve is a three - Lorentzian fit. The highest - frequency peak is
assigned to the totally symmetric pentagonal - pinch A$_{g}$ mode in
isotopically pure $^{12}$C$_{60}$. The other two peaks are assigned to the
perturbed pentagonal - pinch mode in molecules having one and two $^{13}$C -
enriched C$_{60}$, respectively. The insert shows the evolution of these
peaks as the solution is heated. b - the points give the measured
unpolarized raman spectrum in the pentagonal - pinch region for a frozen
solution of $^{13}$C - enriched C$_{60}$ in CS$_{2}$ at 30 K. The solid line
is a theoretical spectrum computed using the sample's mass spectrum, as
described in the text (after [15]).

Fig. 2. Mirror reflection spectra of crystals: 1 - LiH; 2 - LiH$_{x}$D$%
_{1-x} $; 3 - LiD; at 4.2 K. 4 - source of light without crystal. Spectral
resolution of the instrument is indicated on the diagram (after [12]).

Fig. 3. a - Reflection spectra in the A and B excitonic polaritons region of
Cd$^{nat}$S and Cd$^{34}$S at 1.3K with incident light in the $%
\overrightarrow{\text{E}}$ $\perp $ $\overrightarrow{C}$. The broken
vertical lines connecting peaks indicate measured enrgy shifts reported in
Table 18 of Ref. [14]. In this polarization, the n = 2 and 3 excited \
states of the A exciton, and the n = 2 excited state of the B exciton, can
be observed. b - Polarized photoluminescence spectra in the region of the A$%
_{\text{n = 2}}$ and A$_{\text{n = 3}}$ free exciton recombination lines of
Cd$^{nat}$S and Cd$^{34}$S taken at 1.3 K with the $\overrightarrow{\text{E}}
$ $\perp $ $\overrightarrow{C}$. The broken vertical lines connecting peaks
indicate measured enrgy shifts reported in Table 18 of Ref. [14] (after
[22]).

\bigskip

\textbf{References.}

\bigskip

1. M. Jammer, Concepts of mass in classical and modern physics, Harvard
University Press, Cambridge - Massachsets (1961).

2. A. Einstein, Ann. Phys. (Leipzig) \textbf{20}, 371 (1906).

3. L.D. Landau, E.M. Lifshitz, The classical theory of fields, Pergamon, New
York (1958).

4. L.B. Okun, Physics Today, June 1989; Uspekhi Fiz. Nauk \textbf{158}, 512
(1989) (in Russian).

5. B.L. Ioffe, Uspekhi Fiz. Nauk \textbf{171}, 1273 (2001) (in Russian);
Progr. Part. Nucl. Phys. \textbf{56}, 232 (2006).

6. C.D. Frogatt, Surveys High Energy Physics \textbf{18}, 77 (2003); The
Problem of Mass , ArX:hep - ph/0312220.

7. A. Dobado and A.L. Maroto, Phys. Rev. \textbf{D60}, 104045-9 (1999).

8. J.J. Kelly, Phys. Rev. \textbf{C70}, 068202 (2004).

9. S.R. Beane, K. Originas and M.J. Savage, Nucl. Phys. \textbf{B768}, 38,
(2007).

10. G.A. Miller, A.K. Opper, E.J. Stephenson, Annual Review of Nuclear
Science \textbf{56}, 253 (2006).

11. G.A. Miller, Phys. Rev. Lett. \textbf{99}, 112001 (2007); The Neutron
Negative Central Charge Density: an Inclusive - Exclusive Connection, ArXiv
0806.3977.

12. V.G. Plekhanov, Phys. Reports \textbf{410}, 1 (2005).

13. M.Cardona, M.L.W. Thewalt, Rev. Mod. Phys. \textbf{77}, 1173 (2005).

14. V.G. Plekhanov, will be published.

15. J. Menendez and J.B. Page, Vibrational spectroscopy of C$_{60}$, in, M.
Cardona and G. Guntherodt, eds., Light Scattering in Solids VIII, Springer,
Berlin - Heidelberg (2000) (Vol.\textbf{\ 76} in Topics in Applied Physics).

16. V.G. Plekhanov, Isotope effects in solid state physics, Academic Press,
San Diego (2001).

17. A.F. Kapustinsky, L.M. Shamovsky, K.S. Bayushkina, Acta Physicochim.
(USSR)\textbf{\ 7}, 799 (1937).

18. R.J. Elliott, J.AA. Krumhansl, P.L. Leath, Rev. Mod. Phys. \textbf{46},
465 (1974).

19. Y.Onodera and Y. Toyozawa, J. Phys. Soc. Japan \textbf{24}, 341 (1968).

20. M. Zhang, M. Ghieler, T. Ruf, Phys. Rev. \textbf{B57}, 9716 (1998).

21. F.I. Kreingol'd, K.F. Lider, M.B. Shabaeva, Fiz. Tverd. Tela \textbf{26}%
, 3940 (1984) (in Russian).

22. T.A. Meyer, M.L.W. Thewalt and R. Lauck, Phys. Rev. \textbf{B69},
115214-5 (2004).

23.J. Grasser and H. Leutwyller, Phys. Reports \textbf{87}, 77 (1982); H.
Leutwyller, Insights and Puzzles in Light Quark Physics, ZrXiv:hep \ -
ph/070063138.

24. M. Gell-Mann, R.J. Oakes, B. Renner, Phys. Rev. \textbf{175}, 2195
(1968).

25. I. F. Ginzburg, Uspekhi Fiz. Nauk (Moscow) \textbf{179}, 525 (2009) (in
Ruassian).

\end{document}